\documentclass[12pt]{article}
\usepackage{amssymb,amsfonts}
\usepackage[top=2.5cm, bottom=2.5cm, left=2.75cm, right=2.75cm]{geometry} 

\font\bfcal=eusb10 at 12pt
\newcommand{\be}{\begin{equation}}
\newcommand{\ee}{\end{equation}}
\newcommand{\beqs}{\begin{eqnarray}}
\newcommand{\eeqs}{\end{eqnarray}}
\def\({\left(}
\def\){\right)}

\def\Lag{ {\cal L}}
\def\DD{ {\hbox{\bfcal D}}}
\def\Dl{ {\cal D}}

\def\N{ {\cal N}}
\def\Z{ {\cal Z}}

\def\vv{{\bf V}}
\def\WW{{\bf W}}

\def\Tr{{\rm Tr}}
\def\mxth{\mathsurround=0pt }
\def\xversim#1#2{\lower2.pt\vbox{\baselineskip0pt \lineskip-.5pt
x  \ialign{$\mxth#1\hfil##\hfil$\crcr#2\crcr\sim\crcr}}}

\newcommand{\bD}{{\bar D}}
\newcommand{\na}{\nabla}
\newcommand{\bna}{{\bar\nabla}}
\newcommand{\ad}{{\dot\alpha}}
\newcommand{\bd}{{\dot\beta}}

\newcommand{\half}{{\textstyle{\frac12}}}

\renewcommand{\a}{\alpha}
\renewcommand{\b}{\beta}

\renewcommand{\d}{\delta}
\newcommand{\D}{\Delta}
\newcommand{\pa}{\partial}

\newcommand{\e}{\epsilon}
\newcommand{\z}{\zeta}

\renewcommand{\l}{\lambda}

\newcommand{\pis}{{\pi\kern-1.28ex /}}

\newcommand{\s}{\sigma}

\newcommand{\Y}{\Upsilon}
\newcommand{\bY}{{\bar\Upsilon}}
\newcommand{\tY}{{\tilde\Upsilon}}
\newcommand{\btY}{{\bar{\tilde\Upsilon}}}
\newcommand{\tc}{{\tilde\chi}}
\renewcommand{\o}{\omega}

\newcommand{\h}{\eta}
\newcommand{\bu}{{\bar u}}
\newcommand{\bz}{{\bar z}}
\newcommand{\by}{{\bar y}}
\newcommand{\bpa}{{\bar\partial}}

\newcommand{\Ka}{{K\"ahler}}
\renewcommand{\O}{{\Omega}}
\newcommand{\var}{\varepsilon}
\newcommand{\OO}{ {\cal O}}
\newcommand{\cpl}{$\mathbb{P}^1$ }
\newcommand{\cpln}{$\mathbb{P}^1$}
\newcommand{\eg}{{\it e.g., }}
\newcommand{\ie}{{\it i.e., }}

\newcommand{\ft}[2]{{\textstyle\frac{#1}{#2}}}

\newcommand{\eqn}[1]{(\ref{#1})}

\begin{document}
\setcounter{page}{0}
\thispagestyle{empty}
\begin{titlepage}
\begin{center}
\hfill UUITP--03/06  \\
\hfill YITP--SB--06--07  \\
\vskip 20mm

{\Large {\bf Properties of hyperk\"ahler manifolds \\[3mm] and their twistor spaces}}

\vskip 10mm
{\bf Ulf Lindstr\" om$^{a,},$ and Martin Ro\v{c}ek$^{b}$}
\bigskip

{\small\it
$^a$Department of  Physics and Astronomy,
Uppsala University, \\ Box 803, SE-751 08 Uppsala, Sweden \\
~\\
~\\
$^{b}$C.N. Yang Institute for Theoretical Physics\\
SUNY, Stony Brook, NY 11794-3840, USA\\~\\}

{\tt ulf.lindstrom@fysast.uu.se}\\
{\tt rocek@max2.physics.sunysb.edu}
\vskip 6mm
\end{center}

\vskip .2in

\begin{center} {\bf ABSTRACT } \end{center}
\begin{quotation}\noindent
We describe the relation between supersymmetric
$\s$-models on hyperk\"ahler manifolds, projective
superspace, and twistor space. We review the essential
aspects and present a coherent picture with a number of new
results.
\end{quotation}
\vfill
\flushleft{\today}
\end{titlepage}
\eject
\small
\vspace{1cm}
\tableofcontents{}
\vspace{1cm}
\bigskip\hrule
\normalsize
\section{Introduction and a succinct mathematical summary}
\setcounter{equation}{0}
This paper collects the insights that we have gained over
twenty years of studying supersymmetric $\s$-models and
hyperk\"ahler geometry. Many of our results have appeared
elsewhere, both in our work and in the work of others.
Here we want to present a coherent view of how 
supersymmetry naturally reveals the 
geometric structure; in particular, we are led to
the twistor spaces of hyperk\"ahler manifolds.

Supersymmetric $\s$-models are described
by an action functional for maps from a spacetime
into a target manifold; we focus on the case
when the target
space of the $\s$-models is hyperk\"ahler 
\cite{Alvarez-Gaume:1981hm}.
Supersymmetry is most naturally studied 
by extending the spacetime to a superspace
with fermionic as well as bosonic dimensions.
$N=2$ supersymmetric $\s$-models in four spacetime
dimensions (as well as their dimensional reductions
in three and two dimensions)\footnote{The 
formalism can also be developed in six dimensions 
\cite{Grundberg:1984xr} as well as five dimensions 
\cite{Kuzenko:2006mv}; however, 
the four dimensional formalism is the most
familiar.} are best described in projective superspace\footnote{
The terminology ``projective superspace'' is historic;
we are not actually considering a projective supermanifold.} \cite{Gates:1984nk}, 
\cite{Karlhede:1984vr}, \cite{Grundberg:1984xr},
\cite{Karlhede:1986mg}, \cite{Lindstrom:1987ks}, 
\cite{Buscher:1987uw}, \cite{Lindstrom:1989ne}, 
\cite{Lindstrom:1994mw}, \cite{Lindstrom:1994eh}, 
\cite{Gonzalez-Rey:1997qh}. Projective 
superspace naturally leads to twistor space 
\cite{Penrose:1976js,Salamon,Hitchin:1986ea,Lindstrom:1989ne,Ivanov:1995cy}. 

We begin with a brief mathematical summary of some of our
main results that also serves as an introduction to hyperk\"ahler
geometry.
A hyperk\"ahler space ${\cal M}$ supports three globally
defined integrable complex structures $I,J,K$ obeying the
quaternion algebra: $IJ=-JI=K$, plus cyclic permutations.
Any linear combination of these,  $aI+bJ+cK$ is again a \Ka
\ structure on ${\cal M}$ if $a^2+b^2+c^2=1$, \ie if
$\{a,b,c\}$ lies on a two-sphere $S^2\backsimeq
\mathbb{P}^1$. The Twistor space $\Z$ of  a hyperk\"ahler
space ${\cal M}$ is the product of ${\cal M}$ with this
two-sphere $\Z= {\cal M}\times\mathbb{P}^1$. The two-sphere 
thus parameterizes the complex structures and we choose
projective (inhomogeneous) coordinates $\zeta$ to describe it (in a patch
including the north pole).   A choice of $\zeta$ corresponds
to a choice of a preferred complex structure, \eg $J$. The
corresponding \Ka \ form $\omega^{(1,1)}$ is a $(1,1)$ two-form 
with respect to $J$.  For this choice,
the two remaining independent complex structures $I$ and $K$
can be used to construct the holomorphic and antiholomorphic symplectic two-forms
$\omega^{(2,0)}$ and  $\omega^{(0,2)}$.  These three
two-forms are conveniently combined into \cite{Hitchin:1986ea}
\be\label{omega}
\Omega(\zeta)\equiv\omega^{(2,0)}+\zeta\omega^{(1,1)}
-\zeta^2\omega^{(0,2)}~,
\ee
which is a section of a two-form valued $\OO(2)$ bundle on \cpln.
For the four-dimensional case, the statement that the 
hyperk\"ahler space obeys the Monge-Amp\`{e}re equation,
\be
2\,\omega^{(2,0)}\, \omega^{(0,2)}=(\omega^{(1,1)})^2 ~,
\ee
simply becomes the identity\footnote{For the four dimensional case,
these ideas were found previously in a different context \cite{Chakravarty:1991bt}.}
\be
\Omega^2=0~.
\ee
For higher-dimensional manifolds the corresponding identity
\be
\Omega^{n+1}=0
\ee
results in a system of equations constraining the geometry 
to be hyperk\"ahler.

For $\z=0$, we can choose local Darboux coordinates (holomorphic with
respect to the complex structure at the north pole) for $\omega^{2,0}$; as we
smoothly rotate the \cpln~of complex structures, we find
Darboux coordinates $\Y^p(\z)$ and  
$\tilde\Y_p(\z)$ that are regular at $\z=0$:
\be
\Omega(\zeta)=i\,d\Y^p(\z)\, d\tilde\Y_p(\z)
\ee
where $p=1,...,n$ and the (real) dimension of ${\cal M}$ is $4n$,
and the exterior derivative acts only along ${\cal M}$ and {\em not}
along the \cpln.
We introduce the real-structure $\mathfrak{R}$ on \cpl defined 
by complex conjugation composed with the antipodal map. 
From \eqn{omega} we see that the two-form $\Omega$ obeys the reality condition
\be
\label{reality}
\Omega(\zeta)=-\z^2\mathfrak{R}(\Omega(\zeta))~; 
\ee
since 
\be 
\mathfrak{R}(\Y^p(\z))=\bar\Y^p(-\frac 1{\z})
\ee
we have
\be\label{realreal}
i\,d\Y^p(\z)\, d\tilde\Y_p(\z)=i\,\z^2d\bar\Y^p(-\frac 1 {\z})\, 
d\bar{\tilde\Y}_p(-\frac 1 {\z})
\ee
The reality relations (\ref{reality},\ref{realreal}) show that
$\Y$ and $\tilde\Y$ are related to $\bar\Y$ and $\bar{\tilde\Y}$ by a symplectomorphism 
up to the $\z^2$-factor. We introduce a generating function $f(\Y,\bar\Y;\z)$ for 
this twisted symplectomorphism:
\be\label{fgen}
\tilde\Y_p=\z\frac{\pa f }{\pa \Y^p }~, \quad \bar{\tilde\Y}_p
=-\frac 1 {\z}\frac{\pa f }{\pa \bar\Y^p}~;
\ee
then 
\be
i\,d\Y^p\, d\tilde\Y_p
=i\,\z\frac{\pa^2f }{\pa\Y^p\pa\bar\Y^q }d\Y^p\, d\bar\Y^q\equiv i\,\z\pa\bar\pa f~,
\ee
where $\pa$ and $\bar\pa$ are respectively holomorphic and anti-holomorphic
exterior derivatives with respect to the complex structure $J$ at the north pole of the \cpln, and again act only on $\cal M$ and not along the \cpln.
The conditions above imply that $\z\frac{\pa f }{\pa \Y^p }$ is regular at the north pole,
and hence, for a contour encircling  $\z=0$,
\be\label{igez}
\oint\frac{d\z }{2\pi i \z }\z^i\frac{\pa f }{\pa \Y^p }=0~,
\quad i\geqslant 2~,
\ee
as well as the complex conjugate relation. 
As we shall see in subsequent sections, this beautiful mathematics follows from the 
$\s$-model. In particular, \eqn{igez}
are the equations of motion, and $f$ is the projective superspace Lagrangian. 
Thus the function $f$ has the r\^{o}le both of the
superspace $\s$-model Lagrangian and as a generating
function for north-south symplectomorphisms.

One of our new observations generalizes
a result proven in \cite{Hitchin:1986ea} for the special case
when the rotation of the complex structures is generated by an
isometry of the manifold. In general,
rotations of the sphere of complex structures
correspond to nonholomorphic diffeomorphisms on 
the hyperk\"ahler manifold. In twistor
space we can compose such a rotation 
with the corresponding diffeomorphism 
to construct a
symplectomorphism preserving $\Omega$ (up to the $\z$
factor). Going to Darboux-coordinates for
$\omega^{(2,0)}$ we can analyze the effect of these
rotations on the \Ka \ potential $K$. It does not transform
simply under rotations of the complex structures but the net
result is always a new $\tilde K$. We find that for any hyperk\"ahler
manifold, the {\em moment
map} for transformations with respect to rotations about an axis is
the {\em \Ka \ potential} with respect to any complex structure in the
equatorial plane normal to the axis.

\section{Review of projective superspace and SUSY $\s$-models}
\setcounter{equation}{0}
The projective superspace\footnote{A related \cite{Kuzenko:1998xm}
formalism is harmonic 
superspace, as described in \cite{Galperin:2001uw} and references 
therein.} approach to $N=2$ supersymmetry has 
been discussed many times
\cite{Gates:1984nk,Karlhede:1984vr,Hitchin:1986ea,Lindstrom:1987ks,Lindstrom:1989ne}; 
a concise but extensive review
can be found in the appendicies of \cite{dwrv}.
Here we review the aspects relevant to this paper.

We want to emphasize that the requirements of supersymmetry 
in spacetime naturally lead to the constructions that we describe,
and lead us to uncover the geometric structures of the target space.

\subsection{Spinor derivatives}
Superspace is a space with both bosonic and fermionic coordinates;
its essential properties are captured in the algebra of the fermionic derivatives.
The algebra of $N=2$ superspace derivatives in four (spacetime) dimensions is
\be
\{D_{a\a},D_{b\b}\}=\{\bD^a_\ad,\bD^b_\bd\}=0
~,~~\{D_{a\a},\bD^b_\bd\}=i\d^b_a\pa_{a\bd}~,
\ee
where $a,b=1...2$ are isospin indicies and $\a,\b$ and 
$\ad,\bd$ are left and right handed spinor
indicies respectively. Mathematically, the $D$'s are Grassmann odd derivations 
that are sections of the self-dual spin-bundle tensored with an associated $SU(2)$ bundle, 
$S_+\otimes\mathbb{C}^2$, and the $\bar D$'s are sections of 
$S_-\otimes\mathbb{C}^2$.
The superspace derivatives $D_{1\a},\bD^1_\ad$ generate an
$N=1$ subalgebra; we will often decompose representations of the
full $N=2$ algebra in terms of $N=1$ representations.

We may parameterize a \cpl 
of maximal graded abelian subalgebras
as\footnote{In many papers, \eg
\cite{Karlhede:1984vr,Karlhede:1986mg,Lindstrom:1989ne,Lindstrom:1994eh,Gonzalez-Rey:1997qh}, 
the role of $D_1$ and $D_2$ are interchanged. However,
this leads to inconvenient identifications of 
the holomorphic coordinates, and we choose 
conventions compatible with \cite{Ivanov:1995cy}.}
\be\label{projder}
\na_\a(\z)=D_{2\a}+\z D_{1\a}~,~~\bna_\ad(\z)=\bD^1_\ad-\z\bD^2_\ad~,
\ee
where $\z$ is the inhomogeneous coordinate 
on \cpl in a patch around the north pole and 
$\bna_\ad(\z)$ is the conjugate
of $\na_\a(\z)$ with respect to the real structure $\mathfrak{R}$ (complex 
conjugation composed with the antipodal map
on \cpln): 
\be
\bna(\z)\equiv -\z\mathfrak{R}(\na(\z))=-\z\na^*(-\frac {1}\z)~.
\ee
\subsection{Superfields}
Superfields are the generalizations of functions and sections of bundles
to superspace. Superfields in projective superspace are by definition annihilated
by the projective derivatives (\ref{projder}); they differ
by their analytic properties on the \cpl parameterized by $\z$.
The most general superfield that describes a scalar multiplet
is the arctic multiplet $\Y$, which is analytic around
the north pole, and its conjugate antarctic multiplet $\bY$,
which is analytic around the south pole \cite{Lindstrom:1987ks}. 
The conjugate is again defined with respect to the real structure $\mathfrak{R}$.
In some cases, we impose a reality condition on $\Y$. Other useful superfields
are tropical; they may have singularities at both poles, but are regular in a
region where the two coordinate patches overlap. 
These are also usually taken to be real. 

Because the derivatives $\na(\z),\bna(\z)$ all anticommute, we may impose the conditions 
\be\label{pcon}
\na_\a(\z)\Y(\z)=\bar\na_\ad(\z)\Y(\z)=0~;
\ee
these imply
\be
\label{upscon}
D_{1\a}\Y_{i-1}+D_{2\a}\Y_i=\bar D^2_\ad\Y_{i-1}-\bar D^1_\ad\Y_i=0~,
\ee
where
\be\label{Yi}
\Y=\sum_{i=0}\Y_i\z^i~.
\ee
The relations (\ref{upscon}) imply the constraints
\be\label{ups01}
\bar D^1_\ad\Y_0=\bar D^{1\ad}\bar D^1_\ad\Y_1=0~.
\ee
If we decompose $\Y$ into its $N=1$ content, 
we see that only the coefficients $\Y_0,\Y_1$
(and their complex conjugates) are constrained 
as $N=1$ superfields--the constraints (\ref{upscon}) do not imply
any constraints in $N=1$ superspace for the remaining coefficients.

\subsection{SUSY $\s$-model Lagrangians}
Field theories describing maps from a spacetime into
a target manifold $M$ are called $\s$-models,
and are generally described by a Lagrangian. The fields
map points of spacetime to points of the target $M$.

The projective superspace Lagrange density 
$F$ of a $\s$-model with a real $4D$-dimensional target $M$
is a contour integral on \cpl 
of an unconstrained function $f(\Y^a,\bY^a;\z)$
of the multiplets $\Y^a$, $a=1\dots D$ as well as the coordinate $\z$:
\be\label{F}
F(\Y^a_i,\bY^a_i)=\oint_C\frac{d\z}{2\pi i\z}\,f(\Y^a,\bY^a;\z)~;
\ee
the function $f$ is real with respect to the real structure modulo terms that
do not contribute to the contour integral, and $F$ is real.
For general polar multiplets, since all we know about $\Y,\bY$ is that they
are analytic near the north and south pole respectively, this is a purely formal
expression and the contour $C$ is not yet defined; 
we will see how to make this into a sensible contour integral below. For other
multiplets, the contour depends on $f$ and in known examples turns out to
be essentially unique.

The Lagrangian is, \eg
\be\label{lag}
L=D_1^\a D_{1\a} \, \bar D^{1\ad}\bar D^1_\ad  \, F~;
\ee
because of the constraints (\ref{pcon}), 
the action $\int d^4x\, L$ is invariant under
the full $N=2$ supersymmetry.
\section{Superspace equations of motion}
\setcounter{equation}{0}
The equations that describe the extrema of the action
can be described in superspace. 
Since the $N=2$ Lagrangian is written with an 
$N=1$ measure (\ref{lag}), the equations of motion that follow from varying
with respect to $\Y$ can best be understood by thinking of
the $N=1$ superspace content of the $\z$-expansion of $\Y$.
The constraints (\ref{upscon},\ref{ups01}) for a general polar multiplet imply that as $N=1$
superfields, all the $\Y_i,~i\ge 2$ are unconstrained. The equations
that follow from varying them are (we suppress the index $a$ that labels
the various $\Y$ superfields):
\be\label{dF}
\frac{\pa F}{\pa\Y_i}=
\oint_C\frac{d\z}{2\pi i\z}\,\z^i
\left(\frac\pa{\pa\Y}f(\Y,\bY;\z)\!\right)=0~~,~~~i\ge 2~.
\ee
Here the contour should really be interpreted as enclosing
$\z=0$; the auxiliary $N=1$ superfields $\Y_{i>1}$ are eliminated in such
a way as to make sense of this contour.
The equations that follow from varying 
with respect to the constrained $N=1$ superfields 
$\Y_1$ and $\Y_0$ can be found by applying 
$\bar D^2_\ad$ and $\bar D^{2\ad}\bar D^2_\ad$ to the $\Y_2$ equation and
then using the $N=2$ constraint (\ref{pcon}) $\bar\na_\ad\frac{\pa f}{\pa\Y}=0$
to re-express the equations in terms of 
$\bar D^1_\ad$ and $\bar D^{1\ad}\bar D^1_\ad$, respectively. 

It is important to distinguish $N=1$ and $N=2$ on-shell constraints. When
the conditions (\ref{dF}) are interpreted in $N=1$ superspace, they serve
only to eliminate unconstrained (auxiliary) $N=1$ superfields, and so they do not put the
$N=1$ theory on-shell. When we impose $N=2$ supersymmetry as described
in the previous paragraph, field equations for the physical $N=1$ superfields follow
from (\ref{dF}), and the theory is fully on-shell.

The equations (\ref{dF}) simply imply that $\frac\pa{\pa\Y}f(\Y,\bY;\z)$
and hence $\pa f\equiv\frac\pa{\pa\Y}f(\Y,\bY;\z) d\Y$
have at most simple poles; here $\pa$ is a holomorphic derivative 
{\em without} a term $d\z\pa_\z$ along $\mathbb{P}^1$ and $d\Y\equiv
\sum \z^i d\Y_i$.
Thus when one imposes the equations (\ref{dF}), 
\be\label{upstild}
\tY\equiv\z\frac\pa{\pa\Y}f(\Y,\bY;\z)
\ee
is again an arctic multiplet.

The conjugate equation
\be\label{bdF}
\frac{\pa F}{\pa\bY_i}=
\oint_C\frac{d\z}{2\pi i\z}\,(-\z)^{-i}
\left(\frac\pa{\pa\bY}f(\Y,\bY;\z)\!\right)=0~~,~~~i\ge 2~.
\ee
similarly implies that $\bpa f(\Y,\bY;\z)$ has at most
simple zeros. Formally, the equations (\ref{dF},\ref{bdF}) can be used to
eliminate the components $\Y_i,\bY_i,i\ge 2$ in terms 
of $\Y_0,\Y_1,\bY_0,\bY_1$. Given such a solution, $\Y$ and $\bY$ become 
maps on $\mathbb{P}^1$; substituting back
into (\ref{F}), for a contour that encloses the relevant singularities, the formal expression now
becomes well defined. In $N=1$ superspace, the equations (\ref{dF},\ref{bdF})
serve to eliminate the $N=2$ superfields that are unconstrained as $N=1$ superfields;
thus the Lagrangian (\ref{F}) results in a well defined $N=1$ superspace action
for the $N=1$ superfields $\{\Y_0,\Y_1,\bY_0,\bY_1\}$, or equivalently, for
the $N=1$ (anti)chiral superfields $\{\Y_0,\tY_0,\bY_0,\btY_0\}$.

\section{The $N=1$ superspace Lagrangian}
\setcounter{equation}{0}
In $N=1$ superspace, the $\s$-model superspace Lagrangian is
the K\"ahler potential expressed as a function of chiral superfields that
geometrically are identified as holomorphic coordinates. Here
we find the $N=1$ superspace Lagrangian that arises after 
solving the equations (\ref{dF},\ref{bdF});
the K\"ahler potential can be written in terms of the 
$N=1$ (anti)chiral superfields $\{z\equiv\Y_0,u\equiv\tY_0,
\bz\equiv\bY_0,\bu\equiv\btY_0\}$:
\be
K(z,\bz,u,\bu)=\oint_C\frac{d\z}{2\pi i\z}\, f - 
u \oint_{O_N}\frac{d\z}{2\pi i\z}\frac1\z\Y- 
\bu\oint_{O_S}\frac{d\z}{2\pi i\z} (-\z)\bY
\ee
where $O_{N,S}$ are the contours around the north and south poles; we can write
\beqs\label{uz}
u=\oint\frac{d\z}{2\pi i\z}\,\tY&,~&\bu = 
\oint\frac{d\z}{2\pi i\z}\,\btY~,\nonumber \\
z=\oint\frac{d\z}{2\pi i\z}\,\Y&,~&\bz 
= \oint\frac{d\z}{2\pi i\z}\,\bY~.
\eeqs

\section{The $2$-form $\O$ and the meaning of the Lagrangian}
\setcounter{equation}{0}
In this section we construct a $2$-form that leads us to a geometric
interpretation of the $N=2$ superspace Lagrangian. As we shall see
in subsequent sections, this $2$-form captures the essential aspects
of hyperk\"ahler geometry.

An essential observation is that (\ref{dF},\ref{bdF}) imply that
\be\label{sigomups}
\O\equiv i\z\pa\bpa f=i\z\frac{\pa^2}{\pa\Y^a\pa\bY^{\bar b}}
f(\Y,\bY;\z)\, d\Y^a d\bY^{\bar b}
\ee
is a section of an $\OO(2)$ bundle. The two-form
$\O$ plays a central role in our understanding of the mathematical structure of the
model. It can also be written as
\be\label{sigomtups}
\O=i d \Y d\tY=i\z^2 d\bY d\tilde{\bY}~
\ee
where $\tilde{\bY}=-\frac1\z\frac\pa{\pa\bY}f$.
Note that because $\Y,\tY$ are arctic and
$\bY,\btY$ are antarctic, equation (\ref{sigomtups}) {\em implies} that $\O$ is
a section of an $\OO(2)$ bundle. 

Equation (\ref{sigomtups}) has the form of a twisted symplectomorphism, and therefore
there should exist a generating function for this transformation. Indeed, (\ref{upstild})
and its conjugate allow us to identify the $N=2$ superspace Lagrangian $f(\Y,\bY;\z)$
as this generating function.\footnote{Superspace Lagrangians with the interpretation of
a generating function of a symplectomorphism have also been discovered in the context
of $\s$-models with bihermitian target spaces \cite{Lindstrom:2005zr}.} 

\section{Generalized $\Y\leftrightarrow\Y$ duality transformations}
\label{upsups}
\setcounter{equation}{0}
Dualities of various sorts have been considered extensively in superspace.
A rather trivial kind results in a diffeomorphism on the target manifold.
In projective superspace, one may generate such a diffeomorphism by relaxing
the regularity constraint on $\Y$ and re-imposing it with an arctic Lagrange multiplier $\tY$:
\be
f(\Y,\bY;\z)\to f(Y,\bar Y;\z)-\frac{\tY Y}\z+\btY\bar Y \z~;
\ee
integrating out $\tY,\btY$ imposes the constraints that $Y,\bar Y$ are arctic
and antarctic respectively; integrating out $Y,\bar Y$ gives a dual Lagrangian
$f(\tY,\btY;\z)$ which is the Legendre transform of $f$. This corresponds to 
simply interchanging the roles of $\Y$ and $\tY$ above.

The interpretation of the $N=2$ superspace Lagrange density $f$ as the generating
function of a twisted symplectomorphism from holomorphic coordinates
adapted to the complex structure at the north pole to those at the south pole allows
us to generalize this duality.

We can construct holomorphic 
symplectomorphisms of $\Y,\tY \to \chi,\tc$  and 
compose them with $f$ to find the transformed 
$N=2$ superspace Lagrange densities. Explicitly, we
consider a generating function $g(\Y,\chi;\z)$ such that
\be
\label{tcg}
\tY=\frac{\pa g}{\pa\Y}~, ~~\tc=-\frac{\pa g}{\pa\chi}~,
\ee
where the explicit $\z$ dependence of 
$g$ is such that $\Y,\tY,\chi,\tc$ are all arctic.
By polar conjugation we have 
$\bar g(\bY,\bar\chi;\ft{-1}\z)$ such that
\be
\btY=\frac{\pa \bar g}{\pa\bY}
~, ~~\bar\tc=-\frac{\pa \bar g}{\pa\bar\chi}~.
\ee
Then the transformed Lagrange density $h(\chi,\bar\chi;\z)$ is given by
\be
\label{hcbc}
h=f(\Y(\chi,\bar\chi;\z),\bY(\chi,\bar\chi;\ft{-1}\z);\z)
+\frac1\z g(\Y(\chi,\bar\chi;\z),\chi;\z)
-\z\bar g(\bY(\chi,\bar\chi;\ft{-1}\z),\bar\chi;\ft{-1}\z)
\ee
where $\Y(\chi,\bar\chi;\z),\bY(\chi,\bar\chi;\ft{-1}\z)$ are determined by
\be
\label{yby}
\frac{\pa g(\Y,\chi;\z)}{\pa \Y}=-\z\frac{\pa f(\Y,\bY;\z)}{\pa \Y}~,~~
\frac{\pa\bar g(\bY,\bar\chi;\ft{-1}\z)}{\pa \bY}
=\frac1\z\frac{\pa f(\Y,\bY;\z)}{\pa\bY}~.
\ee
To check this, we need to see that 
\be
\label{tch}
\tc=-\z\frac{\pa h}{\pa \chi}~;
\ee
using \eqn{hcbc},
we have:
\be
-\z\frac{\pa h}{\pa \chi}=-\z\left(\frac{\pa f}{\pa \Y}\frac{\pa \Y}{\pa\chi}
+\frac{\pa f}{\pa \bY}\frac{\pa \bY}{\pa\chi}\right)
-\frac{\pa g}{\pa\Y}\frac{\pa\Y}{\pa\chi}-\frac{\pa g}{\pa\chi}
+\z^2\frac{\pa \bar g}{\pa\bY}\frac{\pa\bY}{\pa\chi}~;
\ee
from \eqn{yby}, this gives $-\z\frac{\pa h}{\pa \chi}
=-\frac{\pa g}{\pa\chi}$, and hence, from \eqn{tcg},
we find \eqn{tch}. 
\section{$\OO(2n)$-multiplets and Killing spinors}
\label{oomults}
\setcounter{equation}{0}

In this section, we consider projective superfields that
are sections of certain bundles on the \cpln. 
In particular, $\Y\equiv \h_{(2n)}$ may be a section of a
$\OO(2n)$ bundle\footnote{The $\OO(2)$ case is
special because it arises for hyperk\"ahler manifolds
admitting a triholomorphic torus action, and has been
discussed extensively \cite{Karlhede:1984vr,Hitchin:1986ea}.} 
over \cpl \cite{Lindstrom:1987ks,ketov}:
\be\label{oZn}
\Y(\z)\equiv \h_{(2n)}(\z)=(-)^n\z^{2n}\bY(-\frac1\z)~.
\ee
Thus $\h_{(2n)}(\z)$ is a polynomial of order $2n$ in $\z$.
We show that $\s$-models described in 
terms of these $\OO(2n)$-multiplets admit certain local Killing spinors.
These multiplets as well as other special multiplets were 
considered in  \cite{Lindstrom:1987ks}.

\subsection{Supersymmetric $\s$-models and $\OO(2n)$-multiplets}
We begin with a review of $\OO(2n)$-multiplets and the generalized Legendre transform
construction \cite{Lindstrom:1987ks}.

The formal expression for the $\s$-model Lagrangian (\ref{F}) can be made well-defined
without imposing the conditions (\ref{dF},\ref{bdF}) if we impose certain constraints
on $\Y$. Here we focus on the constraint that $\Y$ is a section of an  $\OO(2n)$-bundle.
We may then impose the reality condition (\ref{oZn}):
\be
\Y(\z)\equiv \h_{(2n)}(\z)=(-)^n\z^{2n}\bY(\ft{-1}\z)~;
\ee
$\h_{(2n)}(\z)=\sum_{i=0}^{2n}\z^i\h_i$ is a polynomial of order $2n$ in $\z$ obeying
the constraints:
\be\label{etareal}
\bar\h_i=(-1)^{n-i}\h_{2n-i}~.
\ee
Now we can find a suitable contour (see, \eg the discussion
in \cite{Houghton:1999hr}) and compute the Lagrange density
\be\label{etaF}
F(\h_i)=\oint_C\frac{d\z}{2\pi i\z}\,f(\h;\z)~;
\ee
As for the polar case, the K\"ahler potential is found by eliminating the $N=1$
auxiliary superfields $\h_i,2\le i\le2(n-1)$ and performing a complex Legendre
transform with respect to $\h_1$ and $\h_{2n-1}=(-1)^n\bar\h_1$:
\be\label{etaK}
K(z,\bz,u,\bu)=
F\Big(\h_i(z,\bz,u,\bu)\Big)-
u\,\h_1(z,\bz,u,\bu)-\bu\,\bar \h_1(z,\bz,u,\bu)~,
\ee
where $\h_i(z,\bz,u,\bu)$ are found by solving (preserving the reality conditions
(\ref{etareal})):
\be\label{etaFcon}
z=\h_0~,~~u=\frac{\pa F(\h_i)}{\pa \h_1}~,~~
\frac{\pa F(\h_i)}{\pa \h_j}=0~,~~2\le j\le2(n-1)~.
\ee

\subsection{Four-dimensional hyperk\"ahler manifolds}
We begin by considering 4(real)-dimensional manifolds; the generalization 
to higher dimensions is given later. We prove that a $\s$-model
model description in terms of a $\OO(2n)$-multiplet is possible if
and only if the manifold admits a $2n$-index Killing spinor\footnote{This
was shown using different techniques in \cite{Bielawski:2000tq}.}.

The metric of a 
hyperk\"ahler manifold satisfies the Monge-Amp\`ere equation; 
we can always find holomorphic coordinates such that this has the form
\be
K_{u\bu}K_{z\bz}-K_{u\bz}K_{z\bu}=1~.
\label{maeq}
\ee
This implies that we can write the line element as
\be
ds^2=|k dz|^2+|k^{-1}du+kK_{z\bu}dz|^2
\ee
where 
\be
k\equiv K_{u\bu}^{-\frac12}~.
\ee
We choose frames $\hat e^{A\dot B}$ (here $A,\dot B$ are {\em target} space
spinor indicies)
\beqs
\hat e^{+\dot+}=k d\bz &~,~~ &
\hat e^{+\dot-}=k\bpa K_u~~=~\frac{d\bu}k+ k K_{u\bz} d\bz ~, \\
\hat e^{-\dot-}=kd z &~,~~& 
\hat e^{-\dot+}=-k\pa K_\bu=-\left(\frac{d u}k+ k K_{z\bu}dz\right) ~,
\eeqs
(so that $ds^2=\hat e^{+\dot+}\hat e^{-\dot-}-\hat e^{+\dot-}\hat e^{-\dot+}$).
We compute the connection; it is self-dual, with $\omega^A{}_B=0$;
the nonvanishing terms are
\be
\omega^{\dot +}{}_{\dot+}=-\omega^{\dot-}{}_{\dot-}
=(\bpa-\pa)\ln(k)~,~~
\omega^{\dot+}{}_{\dot-}=K_{u\bu}\bpa
\left(\frac{K_{z\bu}}{K_{u\bu}}\right)~,~~
\omega^{\dot-}{}_{\dot+}=-K_{u\bu}
\pa\left(\frac{K_{u\bz}}{K_{u\bu}}\right)
\ee
The dual vector fields are 
\be
e^{\dot-}{}_+=-k^{-1}\pa_\bz+kK_{u\bz}\pa_\bu
~, ~~e^{\dot+}{}_+=k\pa_\bu~,~~
e^{\dot+}{}_-=k^{-1}\pa_z-kK_{z\bu}\pa_u~,
~~e^{\dot -}{}_-=k\pa_u
\ee
We now construct a rank $2n$ Killing spinor for an
$\OO(2n)$ multiplet $\h$.  The components of $\h$ 
are related to the components of the spinor by:
\be
\h_i=
\left(\!\begin{array}{c} {2n} \\  i  \end{array}\!\right)
\h{}_{\underbrace{+\dots+}_{2n-i}\underbrace{-\dots-}_i}~~,~~~~
\h\equiv\sum_0^{2n} \h_i\z^i
\ee
The Killing spinor equation is
\be
e^{\dot A}{}_{(A}\h_{B_1\dots B_{2n})}=0~
\ee
because we work in a frame where the connection 
1-form $\omega^A{}_B$ vanishes, or
\be
e^{\dot A}{}_-\h_{i-1}+e^{\dot A}{}_+\h_i=0~.
\label{kseq}
\ee
We begin by checking $i=0,1$. In the generalized 
Legendre transform construction above, 
we identify\footnote{In \cite{Lindstrom:1987ks} 
and many other references, the role of $z,u$ is interchanged with
$\bz,\bu$; also, in some references, the $\h $'s are defined 
with an extra overall factor $\z ^{-n}$.}
\be
\h_0=z~,~~\h_1=-K_u~,~~\h_{2n}=(-1)^n\bz~,~~\h_{2n-1}=(-1)^nK_\bu~.
\ee
Then \eqn{kseq} is trivially satisfied for $i=0$. For $i=1$, we have:
\be
e^{\dot+}{}_-z-e^{\dot+}{}_+K_u=k^{-1}-kK_{u\bu}=0~,~~
e^{\dot-}{}_-z-e^{\dot-}{}_+K_u= 
0-k^{-1}K_{u\bz}+kK_{u\bz}K_{u\bu}=0~.
\ee
The $i=2n,2n+1$ equations are just the complex 
conjugates of the above. For $1<i<2n$, we find 
equations that do not have a simple expression in 
terms of the \Ka-potential; however, we can easily 
prove that they are satisfied by studying the 
superspace description of the $\OO(2n)$ 
multiplet $\h$. The superspace constraints (\ref{upscon}) can be written as 
\be
D^\a_1\h_{i-1}+D^\a_2\h_i=0~,
\label{o2neq}
\ee
where $D^\a_a$ are the superspace spinor derivatives 
with isospin indicies $a$ and spinor indices $\a$. 
Note the similarity to \eqn{kseq}. For $i=0,1,2n,2n+1$, \eqn{o2neq} 
is a set of relations between $D^{\a}_a x^\mu$, 
where $x^\mu=\{z,u,\bz,\bu\}$.  Note that these 
relations are exactly the same as those obeyed by 
$e^{\dot\a}_\pm x^\mu$ as a consequence of \eqn{kseq}.
In superspace, however, \eqn{o2neq} is imposed 
as a constraint that defines $\h$. When we eliminate 
the $N=1$ auxiliary superfields $\h_i,1<i<2n-1$, 
and the Legendre transform variables $\h_1,\h_{2n-1}$, 
we must consider $\h_i(x^\mu)$. Then the equations \eqn{o2neq} become:
\be
\pa_\mu\h_{i-1}D^{\a}_1x^\mu+\pa_\mu\h_iD^{\a}_2x^\mu=0~.
\ee
However, since the linear relations between the 
$D^{\a}_a x^\mu$ and the $e^{\dot A}_\pm x^\mu$ 
are the same, this implies relations between 
the $\pa_\mu\h_{i-1}$ and $\pa_\mu\h_i$ that 
guarantee that the Killing spinor equation \eqn{kseq} is satisfied.

The leading component of the Killing spinors 
discussed here is proportional to a coordinate; 
there is a closely related Killing tensor that
can be constructed out of the spinors which
may be easier to define globally. This is
defined by the components of the derivative 
of the Killing spinor that do not vanish:
\be
X^{\dot A}_{B_1\dots B_{2n-1}}\equiv 
\na^{A\dot A}\h_{AB_1\dots B_{2n-1}}~.
\ee
Because the connection is self-dual, these obey the 
Killing tensor equations \cite{Carter:1977pq}
\be
\na^{B_1}_{\dot A} X^{\dot A}_{B_1\dots B_{2n-1}}=0~,~~
\na_{(B_1}^{(\dot B} X^{\dot A)}_{B_2\dots B_{2n})}=0~.
\ee
For $n=1$, this is the well-known triholomorphic 
Killing vector that characterizes the
$\OO(2)$ geometries \cite{Howe:1985sb}.

\subsection{Higher dimensional hyperk\"ahler manifolds}
For four dimensional hyperk\"ahler manifolds, we were able to 
explicitly relate projective superspace and geometry; bolstered by our
success, we can conjecture geometric results from projective superspace for 
the higher dimensional case: In projective superspace, higher dimensional 
target spaces arise when one 
considers models with more independent superfields. 
Depending on the type of multiplets in the model, 
we will get corresponding Killing spinors and
Killing tensors. 

\section{Properties of twistor space}
\setcounter{equation}{0}

For the reader's convenience, we review the properties of twistor spaces summarized
in section 1.1 and relate them to the geometric structure that projective superspace revealed.

The description of hyperk\"ahler geometry that follows from the projective superspace
formulation of $N=2$ supersymmetric $\s$-models leads to a coherent picture in 
twistor space, where the \cpl of graded abelian subalgebra of the $N=2$ superalgebra is
identified with the \cpl of complex structures on the hyperk\"ahler manifold.
The fundamental object is the $2$-form $\O$ (\ref{sigomups}). In terms of the
hyperk\"ahler structure, it can be written as:
\be\label{ok}
\O=\o^{(2,0)}+\z\o^{(1,1)}-\z^2\o^{(0,2)}~,
\ee
where $\o^{(2,0)}$ is a nondegenerate holomorphic $2$-form and $\o^{(1,1)}$
is the K\"ahler form with respect to the complex structure at the north pole of the 
$\mathbb{P}^1$. One may always choose Darboux coordinates $z,u$ for the holomorphic
symplectic structure $\o^{(2,0)}$; extending these to arbitrary complex structures
parametrized by a point $\z$ on the \cpl lifts $z,u$ to $\Y(\z),\tY(\z)$ and leads us to
write
\be\label{otwist}
\O(\z)=id\Y d\tY~,
\ee
with $\Y(\z),\tY(\z)$ such that $\O(\z)$ is projectively real, and hence a section
of $\OO(2)\otimes\,\O^2(M)$. The reality condition implies the existence of a
twisted symplectomorphism from the north pole to the south pole, and consequently the
existence of the generating function $f(\Y,\bY;\z)$. This in particular {\em proves}
that the projective superspace formalism with polar superfields $\Y,\bY$ is completely
general (at least locally in each patch of the hyperk\"ahler manifold, though we see
no obstruction to patching this together over the whole manifold using the general symplectic
transformations of section 6).

An interesting feature of this way of thinking about hyperk\"ahler geometry is
that it naturally leads to two separate problems: (1) What is $f(\Y,\bY;\z)$? and 
(2) What is $\Y(\z)$? In $N=2$ language, the first is an off-shell problem and the second
is the on-shell problem. It may be possible to solve the off-shell problem for, \eg
$K3$, without solving the on-shell problem. This would still be very interesting,
though it would not yield an explicit metric.

The $2$-form $\O$ also allows us to find the system of partial differential equations
that characterize hyperk\"ahler geometry. For a $4D$-dimensional hyperk\"ahler manifold
$M$, the form (\ref{otwist}) clearly obeys
\be\label{hypereq}
\O^{D+1}=0~.
\ee
For $D=1$, this reduces to the usual Monge-Amp\`ere equation. For higher $D$, this
gives a nice system of equations that implies the Monge-Amp\`ere equation. For example,
for $D=2$, expanding in $\z$, we find
\beqs\label{DisZ}
\o^{(2,0)}((\o^{(1,1)})^2-\o^{(2,0)}\o^{(0,2)})&=&0~,\cr
\o^{(1,1)}((\o^{(1,1)})^2-6\o^{(2,0)}\o^{(0,2)})&=&0~,\cr
\o^{(0,2)}((\o^{(1,1)})^2-\o^{(2,0)}\o^{(0,2)})&=&0~.
\eeqs
This implies the Monge-Amp\`ere equation, which in our conventions for general dimension
$D$ is
\be
(\o^{(1,1)})^{2D}-\left(\!\!\begin{array}{c}2D \\D\end{array}\!\!\right)
(\o^{(2,0)}\o^{(0,2)})^D=0~.
\ee

\section{Rotating the complex structures}
\setcounter{equation}{0}

A crucial role both for the twistor structure and for the supersymmetric
$\s$-models is played by rotations of the \cpl combined with corresponding
rotations of the hyperk\"ahler structure on $M$. 
We consider the 2-form $\O$ with $\o^{(2,0)}$ in Darboux coordinates 
$\o^{(2,0)}=\frac{i}2\e_{ij}dz^idz^j$:
\be\label{Omega}
\O(\z)=idzdu+i\pa\bpa K\z +id\bz d\bu \z^2~,
\ee
where $\pa\bpa K=K_{z\bz}dzd\bz +K_{z\bu}dzd\bu 
+K_{u\bz}dud\bz +K_{u\bu}dud\bu$. 
As described in previous sections of this article,
the form $\O$ is a real section of an $O(2)$ bundle, 
where the real structure is defined
by complex conjugation composed 
with the antipodal map $\bar\z\to-1/\z$, and acts on an
$O(2n)$ section $\h=\sum_{0}^{2n}\z^{i}\h_{i}$ as:
\be
\h(\z)=(-)^n\z^{2n}\bar\h(\ft{-1}\z)~.
\ee
For the $\OO(2)$ case, we have
\be
\h_{0}=-\bar\h_{2},\quad \h_{1}=\bar\h_{1}~.
\ee
\subsection{Rotating $\mathbb{P}^1$}
An $SU(2)$ R-symmetry transformation in 
superspace is generated by M\"obius 
transformations of $\z$, and rotates the 
complex structures on the hyperk\"ahler manifold. If we write
\be
\z'=\frac{a\z+b}{c\z+d}~,
\ee
where $ad-bc=1$ and $\bar d =a,\bar c = -b$ for an 
$SU(2)$ transformation, then for $a=1+i\a,b=\b$, and 
$\a=\bar \a$, the infinitesimal transformation of $\z$ is
\be\label{deltaz}
\d\z=\b+2i\a\z+\bar\b\z^2~.
\ee
An $SU(2)$-transformation is generated by 
\be
\underline{\a}\cdot\underline{ J}\equiv 
\sum_{1}^3\a_{i}J_{i}\equiv \a J_{3}+\half \b J_{-}+\half \bar \b J_{+}~,
\ee
where the $SU(2)$-algebra is
\be
J_{\pm}\equiv J_{1}\pm iJ_{2}~, \quad 
[J_{3},J_{\pm}]=\pm J_{\pm}~,\quad [J_{+},J_{-}]=2J_{3}~.
\ee
Writing 
\be
\d\z=[2i\underline{\a}\cdot\underline{ J},\z]
\ee
we may represent the $SU(2)$ generators as
\be
J_-=-i\pa_\z~,~~J_3=\z\pa_\z~,~~J_+=-i\z^2\pa_\z~.
\ee
More generally, we can add a spin piece, and write
\be
J_-=-i\pa_\z~,~~J_3=\z\pa_\z-\half h~,~~J_+=-i\z^2\pa_\z+ih\z~.
\ee

An $\OO(2n)$ multiplet transforms with $h=2n$, and $\O$ 
transforms with $h=2$ (see, \eg \cite{Karlhede:1984vr} 
and \cite{Gonzalez-Rey:1997qh}).
Then, from $\d\O=-2i(\a J_3+\half \b J_-+\half \bar\b J_+)\O$,  we find
\beqs\label{ddudz}
\d(idzdu)&=&2i\a(idzdu)-\b (i\pa\bpa K)~,\\
\label{dddk}
\d( i\pa\bpa K)&=&-2\b(id\bz d\bu)+2\bar\b(idzdu)~.\\
\label{ddbudz}
\d(id\bz d\bu)&=&-2i\a(id\bz d\bu)+\bar\b (i\pa\bpa K)~,
\eeqs
\subsection{Rotating the hyperk\"ahler structure on $M$}
It is easy to find diffeomorphisms on $M$ that satisfy \eqn{ddudz}:
\be
\d z = i\a h z + \b K_u~,~~
\d u = i\a (2-h) u -\b K_z 
\ee
clearly give the correct transformation. Notice the close 
relation to the Legendre transform construction: 
$-K_u\equiv \h_1$, so $\d z = i\a h z -\b \h_1$. This is exactly 
what we would expect from  projective superspace; 
by changing $h$, we get different $\h$ and or $\Y$ multiplets. 
As the $\a$ transformations are holomorphic, $\pa$ and 
$\bpa$ are invariant under them. Naively,
$K$ transforms as $\d_\a K=i\a[h z K_z+(2-h)u K_u]$; 
we can cancel this by simply subtracting this
from the variation of $K$; thus we define 
$\d_\a K=i\a[h z K_z+(2-h)u K_u]+\d_\a' K=0$. This may look odd, 
but as we shall see, it is very necessary and much
more nontrivial below.

Thus we focus on the $\b$ transformations. We write them as
\be\label{betatrans}
\d_{\b} z^i=\b\e^{ij}K_j~,~~~\d_{\b}\bz^i=0~,
\ee
where $\{z^i\}\equiv\{u,z\}$.
Note that here the naive variation of $K$ vanishes:
$\d_{\b} K=\b\e^{ij}K_jK_i=0$.
Consequently, we have:
\be
\d_{\b}(i\pa\bpa K)=i[d(\d_{\b}z^i)\pa_i\bpa K 
+ dz^i(\d_{\b}\pa_i)\bpa K
+\pa d\bz^i(\d_{\b}\bpa_i)K+\pa\bpa(\d_{\b}'K)]~.
\ee
Because $\d_{\b}\bz^i=0$, we have 
$\d_{\b}\pa_i=-(\pa_i\d_{\b} z^j)\pa_j$, {\it etc.}, and we find
\beqs
\d_\b (i\pa\bpa K)&=&i[d(\d_{\b} z^i)\pa_i\bpa 
K - dz^i(\pa_i\d_{\b} z^j)\pa_j\bpa K
-\pa (d\bz^i(\bpa_i \d_{\b} z^j) K_j)+\pa\bpa(\d_{\b}'K)]\cr
&=&i[(\pa\d_{\b} z^i)\pa_i\bpa K +(\bpa\d_{\b} z^i)\pa_i\bpa 
K -(\pa\d_{\b} z^i)\pa_i\bpa K
-\pa ((\bpa \d_{\b} z^i) K_i)+\pa\bpa(\d_{\b}'K)]\cr
&=&i[(\bpa\d_{\b} z^i)\pa_i\bpa K
-\pa ((\bpa \d_{\b} z^i) K_i)+\pa\bpa(\d_{\b}'K)]\cr
&=&i[(\bpa\d_{\b} z^i)\bpa K_i
-\pa ((\bpa \d_{\b} z^i) K_i)+\pa\bpa(\d_{\b}'K)]~.
\eeqs
Now we substitute (\ref{betatrans}):
\be\label{madel}
i(\bpa\d_{\b} z^i)\bpa K_i=i\b\e^{ij}(\bpa K_j)\bpa K_i
=i\b\e^{ij}d\bz^j d\bz^i=-2i\b d\bu d\bz~,
\ee
where we use the quaternionic relation 
$\o^{(1,1)}[\o^{(2,0)}]^{-1}\o^{(1,1)}=-\o^{(0,2)}$. 
Finally, we need to show that all remaining
terms can cancel. In contrast to (\ref{madel}), 
which is a $(2,0)$ form, the remaining terms $i[
-\pa ((\bpa \d_{\b} z^i) K_i)+\pa\bpa(\d_{\b}'K)]$
are both $(1,1)$ forms. We need to show that 
$\pa ((\bpa \d z^i) K_i)$ is both $\pa$ and 
$\bpa$ closed; this is manifest for $\pa$; 
For $\bpa$, we use (\ref{madel}):
\be
\bpa \pa ((\bpa \d_{\b} z^i) K_i) = 
\pa ((\bpa \d_{\b} z^i)\bpa K_i)
=\pa(-2\b d\bu d\bz)=0~.
\ee
Thus there exists a $\d_{\b}'K$ such 
that the total variation $\d_{\b}( i\pa\bpa K)$ 
is given by (\ref{dddk}).
\subsection{The K\"ahler potential is a Hamiltonian}
A remarkable feature allows us to interpret the K\"ahler 
potential $K$ as a Hamiltonian function. The transformation
(\ref{deltaz}) has a fixed point at $\z=\pm i$ for $\a=0,\b=\bar\b$;
then (\ref{ddudz},\ref{dddk},\ref{ddbudz}) imply that $\d_0\equiv
\d_{\a=0,\b=\bar\b}$ preserves 
\be
[\o^{(2,0)}+\o^{(0,2)}]=\ft12[\O(\z=i)+\O(\z=-i)]~;
\ee
Thus $\d_0$ is a symplectomorphism that preserves $\mathbb{R}e(\o^{(2,0)})$,
and hence is generated by a moment map; this moment map is precisely the
$i$ times the K\"ahler potential:
\be
[\o^{(2,0)}+\o^{(0,2)}](\d_0z^i,\,.\,)=idK~.
\ee
This generalizes the observation in \cite{Hitchin:1986ea}
that for manifolds with an isometry that rotates the complex structure,
the K\"ahler potential can be viewed as the moment map of the
rotation with respect to a complex structure preserved by
the rotation; {\it here we do not need an isometry}.

\section{Normal gauge}
\setcounter{equation}{0}
On any K\"ahler manifold, one can define a {\em normal gauge} 
for the K\"ahler potential \cite{thebook}.
In this gauge, one eliminates any purely holomorphic or 
antiholomorphic pieces using K\"ahler
transformations, and uses holomorphic coordinate 
transformations to make the potential as close as possible to flat:
\be
\label{Knorm}
K=z^i\bz^i+\OO(z^2\bz^2)~,
\ee
{\it i.e.}, all terms except for the flat term are at least quadratic in $z$
and quadratic in $\bz$; these
terms are all expressible in terms of the curvature and its derivatives,
and the explicit expression is easily found by direct computation. 
Clearly, normal gauge is unique up to the choice of base point, 
and up to constant $U(2)$ tranformations.

For a Ricci-flat K\"ahler manifold, 
\be
\det{g_{i\bar j}}=f(z)\bar f(\bz)~;
\ee
in normal gauge, $f(z)$ is constant, as follows 
from \eqn{Knorm}, which implies
\be
\left. (\pa_z)^n g_{i\bar j} \right|_{(z=\bz=0)}=
\left. (\pa_{\bz})^n g_{i\bar j} \right|_{(z=\bz=0)} = 0 ~\forall n ~.
\ee

For a hyperk\"ahler manifold (at least for real D=4), 
we have $(\o^1)^2=(\o^2)^2=(\o^3)^2\propto
\det{g_{i\bar j}}$, and hence $\o^{(2,0)}\o^{(0,2)}$ is 
constant. However, since $\o^{(2,0)}$ is holomorphic,
and its magnitude is constant, we conclude that it is in Darboux 
coordinates (up to a constant phase
which can be absorbed by a constant $U(1)$ transformation 
that preserves the normal gauge); thus:
\be
\o^{(2,0)}=i\, dz^1dz^2~.
\ee

\section{Example: the Eguchi-Hansen geometry}
\setcounter{equation}{0}
In this section we derive the Eguchi-Hansen metric using the methods developed
above. This related to the general program of constructing hyperk\"ahler metrics
on cotangent bundles of symmetric spaces using projective superspace methods
\cite{Gates:1998si,Arai:2006qt,Arai:2006gg,araimore},
and indeed can be applied to all of them. Other recent examples in the projective/twistor
formalism include the explicit elliptic examples of \cite{radu} and the explicit
linear deformations of hyperk\"ahler manifolds given in \cite{Alexandrov:2008ds}.

The Eguchi-Hansen metric lives on the cotangent space $\mathbb{P}^1$;
hence we start with the Fubini-Study \Ka\ potential for $\mathbb{P}^1$
and lift it to $\N=2$ superspace:
\be\label{feh}
f=\ln(1+\Y\bY)~.
\ee
The Eguchi-Hansen metric has a triholomorphic $SU(2)$ isometry which
can be realized by $PSU(2)$ transformations of $\Y$. We can therefore
choose a particular form for $\Y$ and reach general points by acting with
the isometry \cite{Arai:2006qt}. In particular, we make the ansatz that when we set 
$z\equiv\Y(0)=0$ then 
\be\label{upszo}
\Y|_{z=0}=y\z
\ee
is a valid point on the manifold. We now act by a $PSU(2)$ transformation 
which we parameterize so as to recover (\ref{upszo}) as well as $z=\Y(0)$:
\be
\Y\to\frac{z+y\z}{1-y\bz\z}~;
\ee
note that this is a triholomorphic $PSU(2)$ transformation that acts on $\Y$,
{\em not} a rotation of the $\mathbb{P}^1$ of complex structures parameterized by
$\z$.
The conjugate is
\be
\bY= \frac{\by-\bz\z}{z\by-\z}~.
\ee
Following the methods described above, to find $\O$ we need to calculate $\tY$:
\be
\tY=\z\frac{\pa f}{\pa\Y}=\frac{\z\bY}{1+\Y\bY}=
\frac{(\by-\bz\z)(1-y\bz\z)}{(1+z\bz)(1-y\by)}~.
\ee
A quick calculation reveals that $i\,d\Y d\tY$ is indeed a section of $\OO(2)$;
the structure is clarified if we introduce the second holomorphic coordinate
\be
u\equiv\tY(0)=\frac\by{(1+z\bz)(1-y\by)}~,
\ee
which implies
\be
y=\frac{2(1+z\bz)\bu}{1+\sqrt{1+4u\bu(1+z\bz)^2}}~.
\ee
This gives the standard $\O$ for the Eguchi-Hansen \Ka\ form:
\be
\Omega_{EH}=i\,d\Y d\tY=i\,dzdu+\z\omega^{(1,1)}_{EH}+i\,\z^2d\bz  d\bu 
\ee
where
\be
\omega^{(1,1)}_{EH}=-i\,\frac{1+z\bz}{\sqrt{1+4u\bu(1+z\bz)^2}}
\left(dud\bu+\frac{dzd\bz}{(1+z\bz)^3}
+(zdu+2udz)(\bz d\bu + 2\bu d\bz)\right).
\ee
This can be made more familiar by the holomorphic symplectomorphism
\be
u=\frac12 u^{'2}~,~~~z=\frac{z'}{u'}
\ee
which gives
\beqs
\omega^{(1,1)}_{EH}&=&-i\,\frac1{\sqrt{1+r^4}}
\left(r^2(du'd\bu'+dz'd\bz')
+\frac1{r^4}(z'du'-u'dz')(\bz' d\bu'-\bu' d\bz')\right),
\nonumber\\ \nonumber\\
r&\equiv&\sqrt{u'\bu'+z\bz'}~.
\eeqs

This calculation reveals an important feature of our approach and the
virtue of using $\O$: we found the \Ka-form without evaluating any
contour integral; in particular, there are no ambiguities about the orientation
of the contour that can arise in a direct evaluation of the superspace Lagrangian.
An example of such issues is given in Appendix B.

\section{Outlook} 
\setcounter{equation}{0}

We have discussed the intimate relation between twistor space and
supersymmetry as manifested in projective superspace. 

Our primary tools are $N=2$ sigma models with hyperk\"ahler target spaces, but gauging them also introduces gauge connections. These were mainly used here to describe quotient constructions and dualities, but may be studied in their on right in projective superspace. This leaves one obvious gap in the description of models: $N=2$ supergravity. To a certain extent this gap is presently being filled (see \cite{Kuzenko:2008ep} and references therein.)

A more immediate extension of the framework presented here is to include quaternionic 
K\"ahler manifolds. Such an extension is presently under way.

We further note that projective superspace has recently been used to study linear perturbations of a class of hyperk\"ahler metric in \cite{Alexandrov:2008ds}, where an extension to quaternionic 
K\"ahler metrics is also advertised. As our description is fully non-linear, a comparison should be fruitful.




\vspace{5mm}
\noindent{\bf{Acknowledgement:}}
\noindent{We thank the 2003,
2004, 2005, 2006, and 2008 Simons Workshops in Physics and Mathematics at
C.N. Yang Institute for Theoretical Physics and the Department
of Mathematics at Stony Brook
for partial support and for a stimulating atmosphere. We are happy to
thank Claude LeBrun, Blaine Lawson and Dennis Sullivan for many useful discussions
over the years as well as Nigel Hitchin, Lionel Mason, David Skinner, and Rikard von Unge
for more recent comments. We are also happy to thank Sergei Kuzenko
for discussions of the example, and Stefan Vandoren for making \cite{Alexandrov:2008ds}
available to us prior to posting it on the arXiv.
MR thanks the Institute for Theoretical Physics
at the University of Amsterdam for hospitality during the spring of 2006. The work of 
UL is supported in part by VR grant 621-2006-3365 and by 
EU grant (Superstring theory) MRTN-2004-512194.
MR is supported in part by
NSF grant no.~PHY-06-53342.}

\appendix
\section{The hyperk\"ahler quotient in projective superspace}
\setcounter{equation}{0}

For completeness we review constructions having to do with
gauge fields, quotients, and dualities in projective superspace.
The hyperk\"ahler quotient construction was discovered in \cite{Lindstrom:1983rt} and
its geometric interpretation was given in \cite{Hitchin:1986ea}. The tools
to describe it in projective superspace were developed in \cite{Lindstrom:1989ne},
and the description was given in \cite{dwrv}, though it has been known to us
for a long time. Here we review it.

\subsection{Isometries}

The polar multiplet $\Y$ has an infinite number of $N=1$ superfields;
consequently, it is difficult to extract the \Ka\ potential except in
special circumstances. On the other hand, the space of polar
multiplets has an algebraic structure: holomorphic functions of arctic 
multiplets are themselves arctic. This
allows for a very direct realization of triholomorphic isometries of the
hyperk\"ahler geometry in projective superspace: they are simply symmetries
of the projective superspace action (\ref{lag}) that are holomorphic
in the arctic multiplets. 

As we explain below, the whole process of gauging
triholomorphic isometries and performing hyperk\"ahler quotients, when
described in terms of polar multiplets in projective superspace is
essentially the same procedure as for \Ka\ quotients described in terms of
chiral superfields in $N=1$ superspace \cite{Hull:1985pq},\cite{Hitchin:1986ea}.

A triholomorphic isometry acts without rotating the complex structures; therefore
it is generated by a holomorphic vector field $X(\Y)$ that has no explicit dependence
on $\z$:
\be
\d\Y=X(\Y)~,~~\d\bY=\bar X(\bY)~.
\ee
When we gauge a symmetry generated by such a vector field, we introduce a local
parameter $\l(\z)$:
\be
\d\Y=\l(\z)X(\Y)~,~~\d\bY=\bar\l(\ft{-1}\z)\bar X(\bY)~;
\ee
to preserve the holomorphic properties of $\Y$, the parameter $\l(\z)$ must itself be an arctic
superfield, and consequently, $\bar\l(\ft{-1}\z)$ must be antarctic. We are thus led to introduce
a real tropical field $\vv=\mathfrak{R}(\vv)$;
it has coefficients for all powers of $\z$ that are
unconstrained as $N=1$ superfields.
It transforms as
\be
\d\vv=i(\bar\l-\l)~.
\ee
This may be generalized to a nonabelian action, where $\vv,\l,\bar\l$ all become matrix
valued; for a finite transformation by an element $g={\rm e}^{\Lag_{i\l X}}$, we have:
\be
\left(\!{\rm e}^\vv\!\right)={\rm e}^{i\bar\l}{\rm e}^\vv{\rm e}^{-i\l}~.
\ee

Having introduced the field $\vv$, we now show how it describes $N=2$
super Yang-Mills theory \cite{Lindstrom:1989ne}.
We split the tropical gauge multiplet factors regular
at the north and south poles:
\be
{\rm e}^\vv={\rm e}^{\vv_-}\, {\rm e}^{\vv_+}\ ,\qquad \vv_+=
\sum_{n=0}^\infty \vv_{+n}\,\z^n\ ,\qquad \vv_-=\bar \vv_+\ .
\ee
Because $\vv$ is an analytic superfield, $\nabla {\rm e}^\vv =0$, and we may
define a gauge-covariant analytic derivative $\DD$
\be\label{covd}
\DD\equiv \nabla + {\rm e}^{-\vv_-}(\nabla {\rm e}^{\vv_-}) =
\nabla -(\nabla {\rm e}^{\vv_+}) {\rm e}^{-\vv_+}\ .
\ee
Comparing powers of $\z$ for both expressions, we conclude that $\DD$ has
only a constant and a linear term (just as $\nabla$), and hence defines
the $N=2$ gauge-covariant derivative (for a more detailed explanation see
\cite{Lindstrom:1989ne}). This structure is precisely the same
as Ward's twistor construction of self-dual Yang-Mills fields \cite{Ward:1977ta}.
Observe that (\ref{covd}) depends crucially on the reality of $\vv$; 

We find the covariantly chiral gauge
field strength $\WW$ by computing
\be
\{\bar\DD_{\ad}(\z),\frac\pa{\pa\z}(\bar\DD_{\bd}(\z))\}\ =\ \var_{\ad\bd}\, \WW\ .
\label{fieldstrength}
\ee
Note that $\WW$ is $\z$ independent.

We focus on the case when we start with a vector space, and quotient by a linear
(or possibly affine) action; this has the virtue that the formal expression (\ref{F})
for the superspace Lagrangian
can be explicitly evaluated. Thus we start with 
\be\label{fvv}
f(\Y,\bY,\vv)=\bY{\rm e}^\vv\Y
\ee
for any compact group acting linearly on the vector space coordinatized by $\Y$.

We define {\em covariantly} analytic polar multiplets
\be
\hat\Y\equiv {\rm e}^{\vv_+}\Y \ , \qquad \hat{\bar\Y} = \bar\Y
{\rm e}^{\vv_-}\ .
\ee
In terms of these, the gauge-invariant Lagrange density \eqn{fvv} is
quadratic; hence, the $\z$ integral
\be\label{Fapp}
F=\oint_C \frac{d\z}{2\pi i \z}\,\bY{\rm e}^\vv\Y
=\oint_C  \frac{d\z}{2\pi i \z}\,\hat{\bar\Y}\cdot\hat\Y
\ee
can be trivially evaluated,
and the auxiliary superfields can be integrated out to get the
gauge-invariant $N=1$ superspace Lagrangian
\be
L_{N=1}=\hat{\bar z}\cdot\hat z-\hat{\bar s}\cdot
\hat s\ ,
\ee
where $\hat z\equiv\hat\Y_0 $ are $N=1$ gauge-covariantly (vector representation) chiral
superfields and $\hat s\equiv\hat\Y_1$ are modified $N=1$ gauge-covariantly complex
linear superfields
\be
\bar\Dl_{\ad}\hat z=0\ ,\qquad
\bar\Dl^2\hat s =\hat W\hat z\ .
\ee
Here $\hat W$ is the $N=1$ covariantly chiral projection
of the $N=2$ field strength $\WW$ \eqn{fieldstrength} in the representation
that acts on $\hat z$ and $\Dl$ is the $N=1$ gauge-covariant derivative.
We can go to chiral representation and replace
$\hat z,\hat s,\hat W$ with ordinary chiral and linear superfields
$z,s,W$ by introducing the $N=1$ gauge potential $V$:
\be
{\rm e}^V\equiv {\rm e}^{V_-}
{\rm e}^{V_+}\ , \ \ \
\hat z = {\rm e}^{V_+} z\ , \ \ \ \hat s = {\rm e}^{V_+} s\ , \ \ \
\hat W = {\rm e}^{V_+} W{\rm e}^{-V_+}\ ,
\ee
where $V_\pm\equiv\vv_{0\pm}$ is the $N=1$ projection of the $\z$-independent coefficients
of $\vv_\pm$. These substitutions lead to
\beqs
L_{N=1}&=&\bar z\, {\rm e}^V  z-
\bar s\, {\rm e}^V s\ ,\ \label{lagn1s} \\ [3mm]
\bar D^2 s &=& W z\ . \label{sconst}
\eeqs
It is convenient to rewrite the $N=1$ Lagrangian \eqn{lagn1s} in terms of
chiral superfields; to do this, we impose the constraints \eqn{sconst} by
chiral Lagrange multipliers $u$ in a superpotential term
\be
u(\bar D^2 s - W z)\ ,
\ee
and integrate out $s$ to obtain the nonabelian generalization of the $N=1$
gauged Lagrangian (after relabeling $z\to z_+,u\to z_-$):
\be
L_{N=1}= \bar z_+\, {\rm e}^V z_+-
z_{-}{\rm e}^{-V}\bar z_-\ . \label{Ln=1-gauged}
\ee
In addition, we are left with a superpotential
term
\be
\Tr \,[\,W \mu^+]=z_-Wz_{+}\ ,
\ee
where $\mu^+$ is just the holomorphic
moment map. Observe that interchanging $z_+\leftrightarrow
z_-$ and changing the representation of $V$ to its conjugate does not modify
the gauged Lagrangian \eqn{Ln=1-gauged}; this implies that in the original
$N=2$ Lagrangian $F$ \eqn{Fapp}, we can take $\Y$ transforming in the
conjugate representation ({\it e.g.}, opposite charge for U(1)) without
changing the final result. This interchange can be implemented directly in
projective superspace by the $\Y\leftrightarrow\Y$ duality transformation
of Section \ref{upsups} ($\tY$ naturally transforms in the conjugate representation
to $\Y$). In the next subsection we integrate out the $N=2$
gauge fields to find the quotient Lagrangian; in $N=1$ superspace,
integrating out the chiral superfield $W$ imposes the moment map constraint
$\mu^+=0$. 

\subsection{Quotients and Duality}

Just as $N=2$ isometries and gauging in projective superspace bear a
striking resemblance to their $N=1$ superspace analogs, so do $N=2$
quotients and duality; indeed, the tensor multiplet projective superspace
Lagrangian is just the Legendre transform of the polar multiplet
Lagrangian.

The procedure we follow is the same as in $N=1$ superspace: we gauge
the relevant isometries as above; to perform a quotient, we simply
integrate out the gauge prepotential ${\rm e}^\vv$. Since this
does not break the isometry, we are left with an action defined on the
quotient space. To find the dual, we add a Lagrange multiplier $\eta$ that
constrains the gauge prepotential to be trivial\footnote{As explained in 
\cite{Hitchin:1986ea,Rocek:1991ps}, this is the correct geometric way of 
understanding duality;
when one chooses coordinates such that the killing vectors generating the
isometries are constant, this gives the usual Legendre transform.}, and
again integrate out $\vv$; the dual field is then the Lagrange multiplier
$\eta$. As in the $N=1$ case, we only consider duality for abelian
isometries. In that case, the Lagrange multiplier term that constrains $\vv$
is
\be
\frac\eta\zeta \vv\ ,
\ee
where $\eta$ is the $O(2)$ superfield that describes the $N=2$ tensor
multiplet \cite{Karlhede:1984vr}.

\section{Dualities and contour ambiguities}
The Eguchi-Hansen metric can also be described in terms of the $\OO(2)$-multiplet 
\cite{Karlhede:1984vr}
(these are particular instances of the multiplets described in Section \ref{oomults}). 
A particularly nice way of finding this description involves the quotient and duality
described in the previous appendix. Starting from \eqn{feh}, one can write 
\be
f_{\vv}=\ln(1+e^\vv)-\frac{\eta_{(2)}}\zeta \vv
\ee
where $\eta_{(2)}$ is an $\OO(2)$-multiplet; eliminating $\eta_{(2)}$ imposes
the condition that $\vv\propto \ln(\Y\bY)$, whereas eliminating $\vv$ gives:
\be\label{feta}
f_\eta=-\frac{\eta_{(2)}}\zeta
\ln\frac{\eta_{(2)}}\zeta-\left(1-\frac{\eta_{(2)}}\zeta\right)\ln
\left(1-\frac{\eta_{(2)}}\zeta\right)~.
\ee
The metric can be found by evaluating the $\zeta$ integral along a contour first given
in \cite{Karlhede:1984vr} with the caveat that the opposite orientation must be used for the
two terms in (\ref{feta}) to obtain a metric with definite signature.

On the other hand, we can rewrite \eqn{feh} in terms of the symplectic conjugate
variables $\tY$:
\be
\tilde f = \ln\left(1+\sqrt{1-4\tY\btY}\,\right)-\sqrt{1-4\tY\btY}~.
\ee
Performing the duality transformation to the $\OO(2)$ multiplet $\eta_{(2)}$
as above, we obtain:
\be\label{fetare}
\tilde f_\eta=-\frac{\eta_{(2)}}\zeta
\ln\frac{\eta_{(2)}}\zeta-\left(1+\frac{\eta_{(2)}}\zeta\right)\ln
\left(1+\frac{\eta_{(2)}}\zeta\right)~.
\ee
The difference in relative sign between the terms in \eqn{feta} and \eqn{fetare} mean that we
need to use different orientations of the contours when evaluating the metric in the two cases.
Clearly the issue of  contours, in particular their orientation, is a subtle one. In the definition of  $\Omega$ no ambiguities exist, as illustrated in Sec. 11. We thus determine the integration contours by requiring agreement with an $\Omega$ derivation. It would be interesting to compare this idea to the discussions of contours presented in \cite{Houghton:1999hr,Alexandrov:2008ds}.

\newpage

\end{document}